\documentclass[aps,prl,showpacs,twocolumn,nofootinbib,preprintnumbers]{revtex4-2}
\usepackage{bbm}
\usepackage{mathrsfs}
\usepackage{epsfig}
\usepackage{soul,xcolor}
\usepackage{graphicx}
\usepackage{amsfonts}
\usepackage{amsthm}
\usepackage[figuresright]{rotating}
\usepackage{amssymb}
\usepackage{amsmath}
\usepackage{dcolumn}
\usepackage{physics}
\usepackage{float}
\usepackage{bm}
\usepackage{physics}
\usepackage{verbatim}
\usepackage{braket}
\usepackage[normalem]{ulem}
\usepackage[ruled,vlined,linesnumbered]{algorithm2e}
\usepackage{setspace}
\usepackage{lipsum}
\usepackage[colorlinks,linkcolor=blue,anchorcolor=blue,citecolor=blue,urlcolor=blue]{hyperref}
\usepackage{booktabs} 
\usepackage{enumitem,kantlipsum}

\newcommand\blfootnote[1]{%
  \begingroup
  \renewcommand\thefootnote{}\footnote{#1}%
  \addtocounter{footnote}{-1}%
  \endgroup
}

\begin{document}

\title{Foundation Neural Effective Hamiltonian for Strongly Correlated Quantum Materials}
\author{Lixing Zhang$^{1\ast}$, Hongjie Jiang$^{2\ast}$, Di Luo$^{3,4 \dagger}$\thanks{email: \url{diluo1000@gmail.com}}}

\affiliation{$^{1}$\mbox{Department of Chemistry and Biochemistry, University of California, Los Angeles, CA 90095, USA}\\
$^{2}$\mbox{School of Mathematical Sciences, Peking University, Beijing 100871, China} \\
$^{3}$\mbox{Department of Physics, Tsinghua University, Beijing 100084, China} \\
$^{4}$\mbox{Institute of Advanced Study, Tsinghua University, Beijing 100084, China} \\
}

\begin{abstract}

Simulating strongly correlated quantum materials often involves not a single Hamiltonian, but a family of Hamiltonians whose ground states evolve across experimentally tunable couplings.
Foundation neural quantum states (FNQS) offer a promising route to amortizing many-body calculations across such families, but can lose accuracy near phase transitions and still incur non-negligible sampling costs that grow with the number of target couplings. We introduce the Foundation Neural Effective Hamiltonian (FNEH), which projects a Hamiltonian family onto a compact subspace spanned by FNQS sampled at selected couplings. By variationally combining FNQS across parameter space, FNEH systematically improves their ground-state approximation and can recover phase boundaries that the foundation model misidentifies. Once the required operator matrix elements are sampled, FNEH enables sweeps over couplings, observables, and phase boundaries at a cost governed by the small effective-Hamiltonian dimension, without repeated neural-network sampling at every target coupling. We demonstrate FNEH in strongly correlated moiré materials, where it accurately resolves competing phases, enables high-resolution multidimensional phase scans, and substantially reduces the computational cost of exploring many target Hamiltonians. The results open a new avenue for studying strongly correlated quantum materials with foundation models.
\end{abstract}

\maketitle
\blfootnote{$^\ast$ These authors contributed equally to this work}
\blfootnote{$^\dagger$ Email: ~\url{diluo@tsinghua.edu.cn}}

\textit{Introduction---.}
Quantum simulation often involves families of Hamiltonians spanning different couplings and phases, rather than a single fixed Hamiltonian.
Such families arise whenever a quantum system depends on multiple physical or experimental parameters, including interaction strength, pressure, strain, doping, external fields, dielectric screening, lattice depth, geometry, or chemical composition~\cite{keimer2017physics,basov2017towards,bloch2008manybody}. Mapping phases, response functions, and optimized materials properties requires solving many nearby strongly interacting Hamiltonians whose ground states may change nonperturbatively.

Neural quantum states (NQS) have emerged as accurate variational ansatzes for quantum many-body systems~\cite{carleo2017solving,glasser2018neural,luo2019backflow,pfau2020ab,pfau2024accurate,gu2025solving,luo2025solving,zhang2025neural, chen2025neural, gao2024neural, sharma2026comparing, abouelkomsan2026topological, zaklama2025attention, foster2025ab, qi2026neural}. In parallel, foundation-model ideas from machine learning suggest that a single large model can capture structure shared across many tasks and then generalize through zero-shot or few-shot adaptation~\cite{brown2020language,jumper2021highly}. For quantum simulation, this motivates the development of foundation neural quantum states (FNQS)~\cite{gao2026excited, zaklama2026large, zaklama2025attention, nazaryan2026qernel, rende2025foundation, gao2022sampling, qi2026neural, foster2025ab}, which are conditioned on Hamiltonian parameters, trained once over a parameter manifold and evaluated at new couplings. Such amortization is particularly attractive for materials simulation. However, since FNQS are continuously conditioned on coupling parameters, their predictions can be fragile near level crossings, competing phases, or sharp changes in correlation patterns. Meanwhile, evaluating multiple target couplings requires repeated Monte Carlo sampling, leading to a sampling cost that grows linearly with the number of couplings.

On the other hand, eigenvector continuation (EC) constructs a compact subspace from eigenstates sampled at selected points in parameter space, allowing nearby states to be reconstructed efficiently~\cite{mejuto2023quantum,frame2018eigenvector,cances2002towards,konig2020eigenvector,yoshida2022constructing,ohlberger2015reduced, duguet2024colloquium}. This subspace viewpoint is particularly appealing near a phase transition, where a single pointwise predictor may fail but states sampled from different regions can collectively represent competing ground-state branches. More broadly, Rayleigh-Ritz, Lanczos, generator-coordinate, and configuration-interaction methods improve variational descriptions by projecting the Hamiltonian onto a compact many-body subspace and solving the resulting eigenvalue problem~\cite{lanczos1950iteration,griffin1957collective}. As conventional basis choices may lose accuracy or become prohibitively large in strongly correlated systems, a key challenge of the approach is to efficiently construct a compact many-body subspace that remains both accurate and scalable.

\begin{figure*}
    \centering
     \includegraphics[width=2\columnwidth]{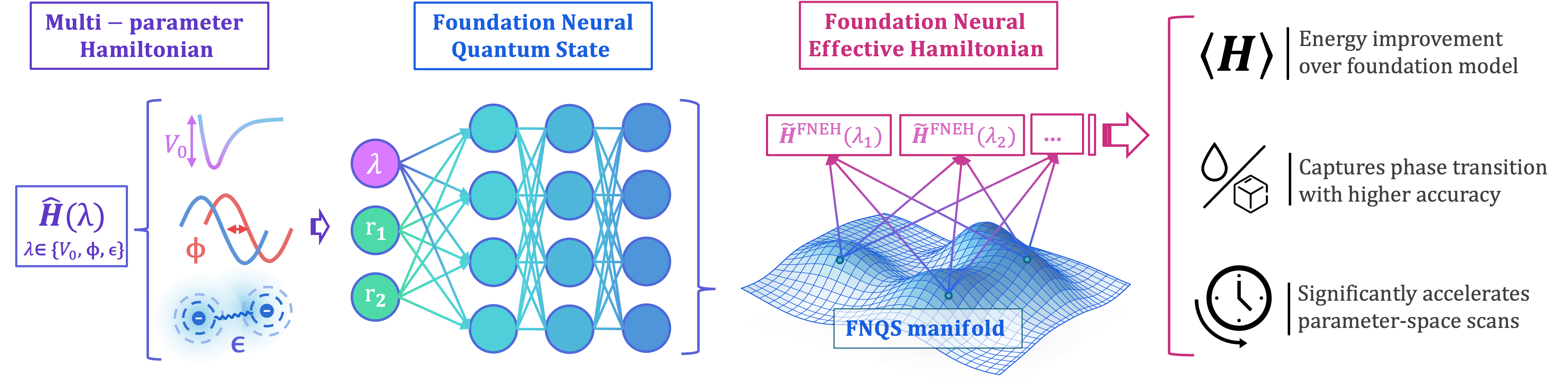}
    \caption{Schematic of the FNEH method. For a multi-parameter Hamiltonian, the parameters $\boldsymbol{\lambda}$ are embedded into an FNQS, which outputs parameter-dependent wavefunction amplitudes. By sampling FNQS wavefunctions at different locations in parameter space, an effective Hamiltonian that depends linearly on $\boldsymbol\lambda$ is constructed. The effective Hamiltonian can then be rapidly reconstructed and solved across parameter space, generating a  manifold of approximate ground-state wavefunctions.}
    \label{fig:schematic}
\end{figure*}

Here we introduce the \emph{Foundation Neural Effective Hamiltonian} (FNEH) for strongly correlated quantum materials. Given a set of foundation neural wavefunctions evaluated at selected Hamiltonian parameters, FNEH projects the full Hamiltonian family into their compact span. First, for a target Hamiltonian, diagonalizing the FNEH variationally improves upon the single foundation-model wavefunction by combining FNQS basis states sampled from multiple nearby couplings. Second, FNEH is constructed directly from pretrained foundation wavefunctions without additional target-specific optimization. Once the overlap and operator matrix elements are sampled, FNEH can sweep through the whole parameter subspace with complexity scaling only with the small number of basis states, enabling rapid phase-transition scans and observable calculations.

We demonstrate FNEH in a continuum model of interacting electrons in a moiré potential, an important platform for studying strongly correlated and topological quantum phases~\cite{bistritzer2011moire,tang2020simulation,regan2020mott,kennes2021moire,wang2024fractional,xia2025superconductivity}. Its broad tunability makes moiré materials a natural testbed for evaluating FNEH across a family of strongly correlated Hamiltonians.

\textit{Foundation Neural Quantum States ---.} For a Hamiltonian specified by parameters \(\boldsymbol{\lambda}=(\lambda_1,\lambda_2,\ldots,\lambda_K)\), an FNQS describes the corresponding wavefunction manifold using a single neural network. Recent studies have developed a variety of FNQS architectures for many-body systems, including lattice systems~\cite{rende2025foundation, zaklama2025attention, qi2026neural}, quantum materials~\cite{zaklama2026large, nazaryan2026qernel}, and molecular systems~\cite{gao2026excited, gao2022sampling, foster2025ab}.

To optimize the FNQS over a region of the parameter space $\Lambda_{\rm train}$, we select a discrete set of training points within $\Lambda_{\rm train}$, which leads to the following loss function \cite{rende2025foundation}:
\begin{eqnarray}
\mathcal{L} = \sum_{\boldsymbol\lambda \in \Lambda_{\mathrm{train}}} \frac{\langle \Psi(\boldsymbol\lambda) | \hat{H}(\boldsymbol\lambda) | \Psi(\boldsymbol\lambda) \rangle}{\langle \Psi(\boldsymbol\lambda) | \Psi(\boldsymbol\lambda)\rangle}
\end{eqnarray}

In this work, we construct the FNQS based on Psiformer~\cite{von2022self} with additional embeddings of the Hamiltonian parameters. As detailed below, our proposed FNEH is compatible with any FNQS and systematically improves upon the underlying FNQS. Recent architectures for FNQS in continuum electronic systems~\cite{zaklama2026large, nazaryan2026qernel} are also applicable under FNEH. In our FNQS, the Hamiltonian parameters are first embedded as normalized features $\widetilde{\mathbf{e}_{\rm h}}$. Together with electron features $\mathbf{e}_{\rm elec}$, which preserve periodic boundary conditions of the system, the features are passed to the transformer backbone of Psiformer. This yields the conditional wavefunction amplitude $\Psi(\boldsymbol \lambda, \boldsymbol{R})$, where $\boldsymbol{R} = (r_1, r_2, \dots r_{N_e})$ denotes the many-body particle positions. Details of the FNQS architecture and optimization procedure are provided in the Appendix.

\textit{Foundation Neural Effective Hamiltonian ---.}
A well-trained FNQS minimizes the energy over the training region. However, FNQS face two crucial challenges: accurately capturing sharp wavefunction changes across phase boundaries and the growing sampling cost associated with evaluating many target couplings. To resolve this issue, we introduce FNEH. Consider a Hamiltonian with $K$ terms of the following form:  
\begin{eqnarray}
{\hat H}(\boldsymbol\lambda) = \sum_k \lambda_k {\hat h}_k
\end{eqnarray}
Define a set of wavefunctions constructed by one FNQS $\{ \ket{\Psi(\boldsymbol\lambda_1)}, \ket{\Psi(\boldsymbol\lambda_2)}, \dots \ket{\Psi(\boldsymbol\lambda_{N_{\rm state}})} \}$, where $N_{\rm state}$ is the total number of basis states. The FNEH method seeks to construct a series of $\boldsymbol\lambda$-free matrices ${\hat A}_k$ such that:
\begin{eqnarray}
{\hat A}_k = {\hat S}^{-1}{\tilde h}_{k}
\end{eqnarray}
where $[{\tilde h}_{k}]_{i,j} = \langle \Psi(\boldsymbol\lambda_i) | \hat{h}_k | \Psi(\boldsymbol\lambda_j) \rangle$, and ${\hat S}_{i,j} = \langle \Psi(\boldsymbol\lambda_i) | \Psi(\boldsymbol\lambda_j) \rangle$. By adding $\boldsymbol\lambda$ back to the equation, the FNEH can be constructed as:
\begin{equation}\label{EQ:linear_FNEH}
 {\hat H}^{\rm FNEH}(\boldsymbol\lambda) = \sum_k \lambda_k  {\hat A}_k
\end{equation}
By saving ${\hat A}_k \in \mathbb{C}^{N_{\rm state} \times N_{\rm state}}$, $\hat H^{\rm FNEH}$ can be reconstructed at every $\boldsymbol \lambda$ without any additional calculations. Diagonalizing $\hat H^{\rm FNEH}(\boldsymbol \lambda)$ provides immediate access to the system energy and the corresponding ground state within the subspace spanned by $\ket{\Psi (\boldsymbol\lambda_i)}$ with optimal coefficients $\mathbf{\tilde c} = (c_1, c_2, \dots)$. 
FNEH combines a foundation-model perspective with the advantages of variational subspace methods, leveraging FNQS to efficiently generate accurate and scalable basis states across a range of couplings, which advances recent efforts on related NQS-based subspace approaches~\cite{hendry2025grassmann, kahn2026variational, pfau2024accurate, li2026efficient}.

The estimation of ${\hat A}_k$ requires Monte Carlo sampling. To construct high-accuracy FNEH, we use the determinant-state sampling (DSS) method~\cite{pfau2024accurate, hendry2025grassmann, kahn2026variational}. Define extended electron configuration $\{{\boldsymbol R}\} = ({\boldsymbol R}_1, \dots, {\boldsymbol R}_{N_{\rm state}})$. The determinant state can be written as:
\begin{equation}
\Psi_{\rm det}(\{{\boldsymbol R}\}) =\det[\boldsymbol{\Psi}]
\end{equation}
where $\boldsymbol{\Psi} \in \mathbb{C}^{N_{\rm state} \times N_{\rm state}}$ with $[\boldsymbol{\Psi}]_{i,j} = \Psi(\boldsymbol\lambda_j,\boldsymbol{R}_i)$. By sampling extended coordinates from $p_{\rm det} = |\Psi_{\rm det}(\{{\boldsymbol R}\})|^2$, we have:
\begin{equation}\label{EQ:det_hamil}
{\hat A}_k = \mathbb{E}_{ \{{\boldsymbol R}\} \sim p_{\rm det}}\left[ \boldsymbol{\Psi}^{-1}[{\hat h}^{(k)}_{\rm det} \boldsymbol{\Psi}]
\right]
\end{equation}
where $[\hat{h}_{\rm det}^{(k)} \boldsymbol{\Psi}]_{i,j} = \hat{h}_k \Psi(\boldsymbol\lambda_j, {\boldsymbol R}_i)$. DSS avoids direct sampling of the overlap matrix \(\hat S\) and the associated amplification of statistical errors during its inversion. This improves the accuracy and stability of the constructed FNEH, enabling a more reliable resolution of the phase transition. A proof of Eq.~\ref{EQ:det_hamil}, together with details of the FNQS architecture and optimization procedure, is provided in the Appendix.

\begin{figure}[b]
    \centering
     \includegraphics[width=1\columnwidth]{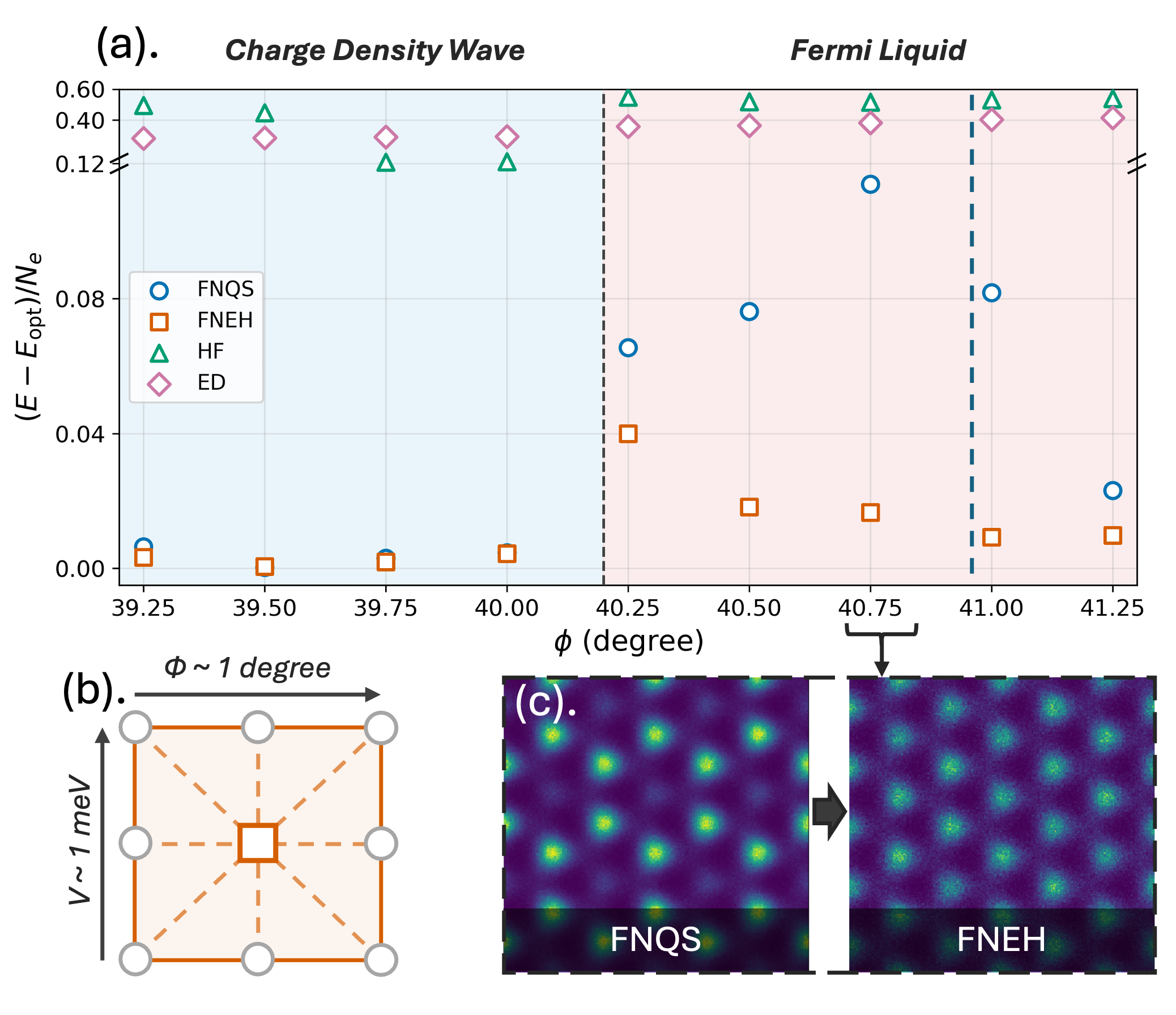}
    \caption{Ground-state energies predicted by different methods. The system parameters are \(N_{\rm site}=27\), \(N_e=18\) (2/3 filling), \(V=15\)meV, and \(\epsilon=10\). (a) Per-particle energy differences relative to the optimal reference for the FNQS, FNEH, Hartree-Fock (HF), and exact diagonalization (ED). Background colors indicate the different phases. The left and right dashed lines indicate the phase boundaries predicted by the optimal reference and FNQS, respectively. (b) Illustration of the basis-state selection used to construct the FNEH. (c) Charge densities predicted by the FNQS and FNEH at \(\phi=40.75^\circ\).}
    \label{fig:energy_bench}
\end{figure}

Compared with FNQS, FNEH is variationally guaranteed to be no worse than any individual FNQS basis state, while also improving sampling efficiency by reusing a compact projected Hamiltonian across target couplings. Since FNQS vary continuously with the conditioning parameters, they may struggle to capture abrupt wavefunction changes across phase transitions. FNEH overcomes this mismatch by variationally combining FNQS basis states from different parameter regions, thereby accommodating discontinuous changes between competing ground states. FNEH is also compatible with any FNQS architecture and can systematically benefit from improved FNQS as their accuracy advances. Compared with conventional Eigenvector Continuation\cite{frame2018eigenvector}, FNEH substantially reduces the cost of constructing the subspace itself. Rather than obtaining each basis state from an independent many-body calculation, single pretrained FNQS generate many-body wavefunction basis across parameter space by inference, amortizing the cost of basis generation over the entire Hamiltonian family. This is particularly advantageous when conventional continuation relies on expensive high-accuracy solvers, such as Exact diagonalization (ED) or full configuration interaction, whose cost rapidly grows with system size. FNEH further integrates this inexpensive basis construction with determinant-state sampling, which efficiently estimates the projected Hamiltonian without separately sampling a noisy overlap matrix. Finally, its operator-resolved form separates the parameter dependence from the sampled matrix elements, allowing the same effective Hamiltonian to be reconstructed and diagonalized throughout parameter space without additional sampling. Together, these features make FNEH substantially scalable for dense many-body parameter scans.

\textit{Results ---.} We demonstrate the FNEH method using a heterobilayer moiré Hamiltonian describing electrons confined to a single semiconductor layer and subject to a moiré potential and Coulomb interactions. The Hamiltonian of the moiré system can be written as\cite{wu2018hubbard}:
\begin{eqnarray}\label{EQ:Main_Hamil}
\hat{H} = \sum_i [-\frac{\nabla_i^2}{2 m^{\ast}} - 2V_0 V(\boldsymbol{r}_i, \phi) ] + \frac{e^2}{2\epsilon}\sum_{\langle i, j \rangle}U(\boldsymbol{r}_i - \boldsymbol{r}_j)
\end{eqnarray}
The above Hamiltonian is parametrized as: $\hat{H} \equiv \hat{H} (V_0, \phi,\epsilon)$, where $V_0$ and $\phi$ are the strength and phase of the moiré potential, respectively. $\epsilon$ is the dielectric constant. $m^\ast$ is the effective mass of the electron, which is fixed at $0.35 m_e$. $V(\boldsymbol{r}_i, \phi) = \sum_{j=1}^3\cos(\boldsymbol{g}_j \cdot \boldsymbol{r} + \phi)$ and $U(\boldsymbol{r}) = \sum_{n,m} 1/{|\boldsymbol{r} + n \boldsymbol{L}_1 + m \boldsymbol{L}_2}|$. The moiré lattice period $a_{M}$ is set as $8.031$nm. $V(\boldsymbol{r}_i, \phi)$ can be further rewritten as $\cos(\phi) V_{\rm c}(\boldsymbol{r}_i) + \sin(\phi) V_{\rm s}(\boldsymbol{r}_i)$, where $V_{\rm c}(\boldsymbol{r}_i) = \sum_{j=1}^3\cos(\boldsymbol{g}_j \cdot \boldsymbol{r})$ and $V_{\rm s}(\boldsymbol{r}_i) = -\sum_{j=1}^3\sin(\boldsymbol{g}_j \cdot \boldsymbol{r})$. Here, $\boldsymbol{g}_i = \frac{4\pi}{\sqrt{3 a_M}}(\cos\frac{2\pi i}{3}, \sin\frac{2\pi i}{3})$ is the reciprocal lattice vector. This allows us to rewrite Eq.~\ref{EQ:Main_Hamil} linearly with respect to scalar functions of the Hamiltonian parameters.

In Fig.~\ref{fig:energy_bench}, we benchmark FNEH against several reference methods. Here, optimal reference (Opt. Ref) denotes the wavefunction obtained by optimizing a separate NQS with same architecture independently at each Hamiltonian parameter, and therefore provides a variationally better reference than the corresponding FNQS prediction. FNQS denotes the wavefunction given by FNQS without any additional optimization. We also compare against Hartree--Fock (HF) and exact diagonalization (ED) results; details of the HF and ED implementations are provided in the Appendix. We first train an FNQS using 12 training points distributed evenly across $V = [14 \text{meV}, 16 \text{meV}]$ and $\phi = [35^{\circ}, 45^{\circ}]$. For the rest of the parameters, we use $N_{\rm site} = 27$ with $N_e = 18$ (2/3 filling) and $\epsilon = 10$. To benchmark FNEH along the $V-\phi$ parameter space, we fix $V = 15$meV and consider target points ranging from $\phi = 39.25^{\circ}$ to $\phi = 41.25^{\circ}$. At each target point, we construct an FNEH using 9 FNQS states in its local neighborhood, with $\Delta V,\Delta\phi\in\{-0.5,0,0.5\}$, and obtain the improved variational energy by diagonalizing the FNEH. Within the FNQS training region, we observe that the optimal reference exhibits a charge density wave (CDW) to Fermi liquid (FL) phase transition at around $\phi = 40.2^{\circ}$. However, this phase transition is wrongly captured by the FNQS, which depicts the same phase transition at around $\phi = 40.9^{\circ}$. This leads to a higher FNQS energy when $\phi > 40.2^{\circ}$. In contrast, FNEH systematically lowers the FNQS energy throughout this region, with a maximum improvement of around $0.1$ meV/$N_e$. More importantly, FNEH recovers the correct phase transition near $\phi = 40.2^{\circ}$. For example, in Fig.~\ref{fig:energy_bench}c, we plot the charge density predicted at $\phi = 40.75^{\circ}$. As the FNQS predicts a spurious CDW state with a dominant crystalline pattern, FNEH corrects that and recovers the FL state. 

\begin{figure}[t]
    \centering
     \includegraphics[width=1\columnwidth]{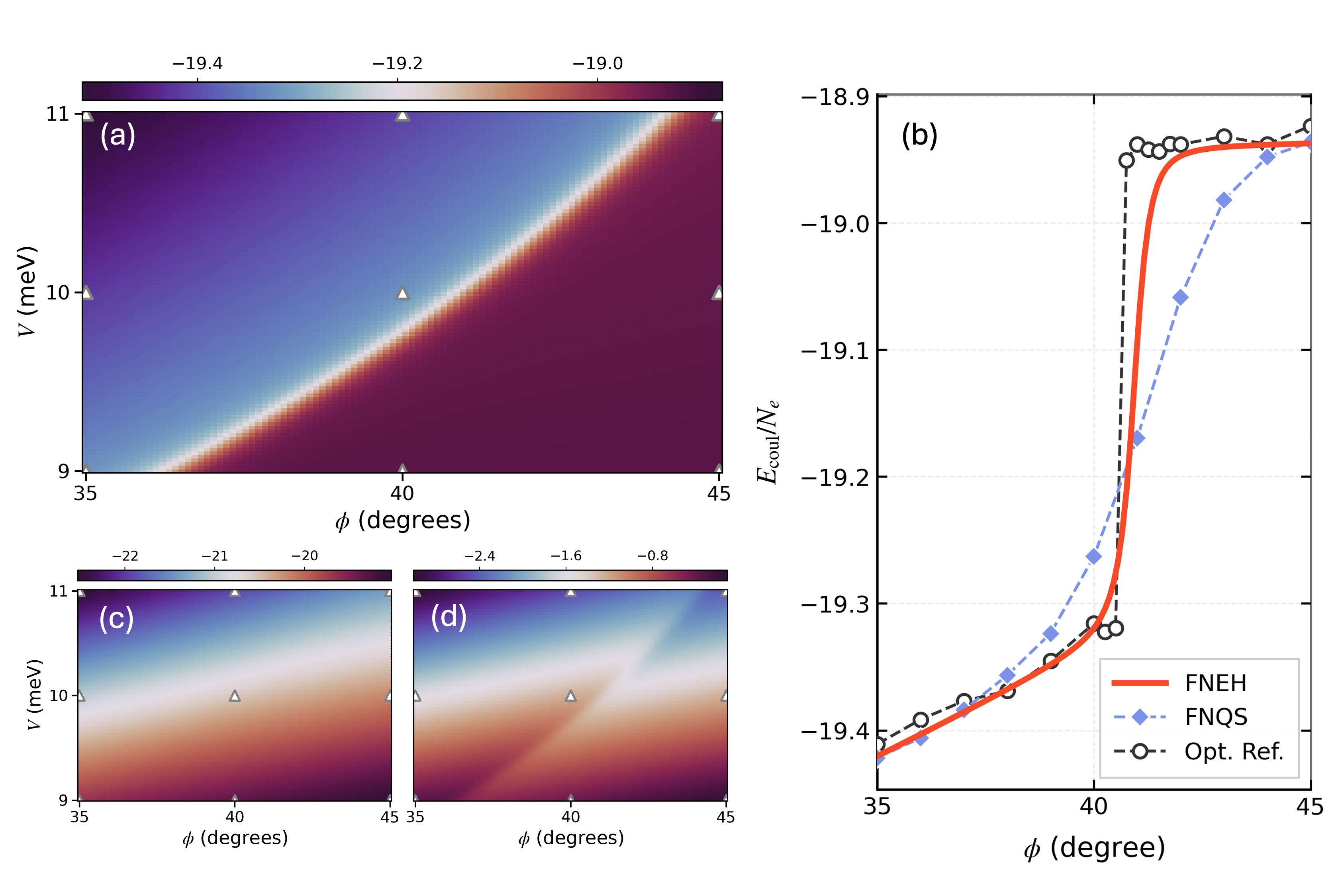}
    \caption{PESs in the $V$--$\phi$ plane predicted by the FNEH. The system parameters are $N_{\rm site}=36$, $N_e=12$ (1/3 filling), and $\epsilon=10$. Gray triangles indicate the basis states used to construct the FNEH. (a) Coulomb-energy PES, $E_{\rm coul}$, predicted by FNEH over $V\in[9\text{meV},11\text{meV}]$ and $\phi\in[35^\circ,45^\circ]$. (b) Cross-section of (a) along the $\phi$ direction at fixed $V=10$meV, compared with the FNQS and optimal reference energies evaluated at $1^\circ$ intervals. (c,d) FNEH-predicted PESs of the total energy, $E_{\rm tot}$, and kinetic energy, $E_{\rm kin}$, respectively.}
    \label{fig:2d_phase_scan}
\end{figure}

In Fig.~\ref{fig:2d_phase_scan}, we demonstrate another important application of FNEH: rapid generation of high-quality wavefunctions across a multidimensional phase boundary. For a system with $N_{\rm site} = 36$, $N_e = 12$ (1/3 filling) and $\epsilon = 10$, we train an FNQS using 9 training points on the grid $V = \{9\text{meV}, 10\text{meV}, 11\text{meV} \}$ and $\phi = \{35^{\circ}, 40^{\circ}, 45^{\circ} \}$. Using the same training points as the basis states (labelled in gray triangles), we construct an FNEH following Eq.~\ref{EQ:linear_FNEH}. By varying the coefficients $\lambda_k = \{-2 V_0\cos(\phi), -2 V_0\sin(\phi)\}$ in  Eq.~\ref{EQ:linear_FNEH}, we solve the FNEH on a $100 \times 100$ grid spanning the FNQS training region. In Fig.~\ref{fig:2d_phase_scan}a, we plot the potential energy surface(PES) of the Coulomb energy $E_{\rm coul} = \langle \frac{e^2}{2\epsilon}\sum_{\langle i, j \rangle}U(\boldsymbol{r}_i - \boldsymbol{r}_j) \rangle$ of the FNEH state. A sharp CDW-to-FL transition is observed across the diagonal of the Coulomb PES predicted by the FNEH. This behavior is expected: the CDW phase lowers the Coulomb energy by forming crystalline charge patterns, whereas the FL phase lowers the kinetic energy, which is dominated by moiré potential. In Fig.~\ref{fig:2d_phase_scan}b, we plot a cross-section of the Coulomb-energy PES predicted by the FNQS, FNEH, and optimal references along the $\phi$ direction at $V=10$meV. Both FNEH and the optimal reference predict the same phase transition near $\phi=40.5^\circ$, characterized by an abrupt change in the energy gradient. In contrast, the PES predicted by the FNQS exhibits a substantially weaker change in slope. This demonstrates that FNEH improves upon the FNQS in resolving the phase transition. In Fig.~\ref{fig:2d_phase_scan}c and d, we plot the total-energy and kinetic-energy PESs, $E_{\rm tot}$ and $E_{\rm kin}$, respectively, with $E_{\rm tot}=E_{\rm kin}+E_{\rm coul}$. We observe that the total-energy PES remains smooth across the transition as expected.


\begin{figure}[t]
    \centering
     \includegraphics[width=1\columnwidth]{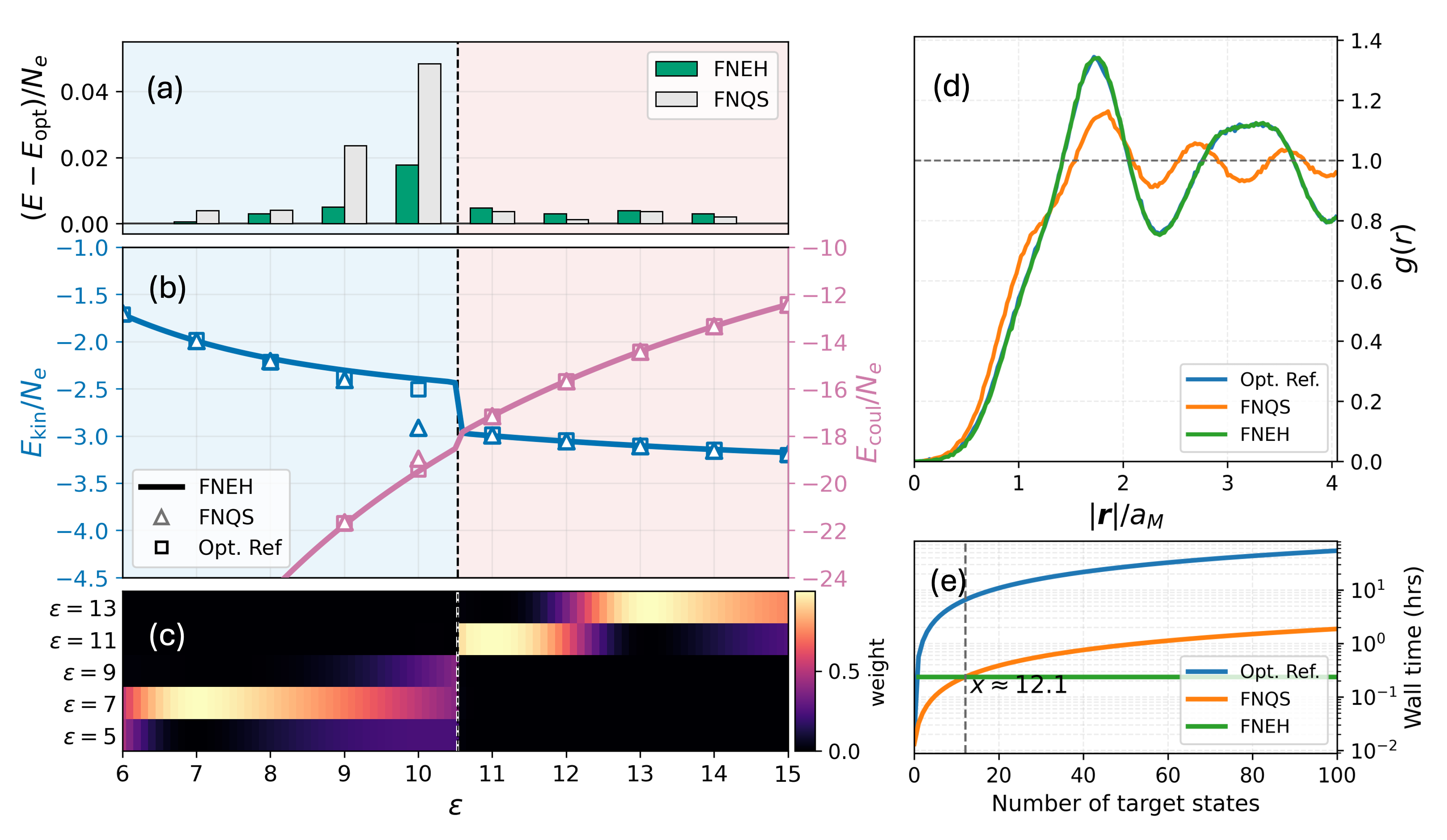}
    \caption{FNEH captures the phase transition across $\epsilon$ while demonstrating high computational efficiency. (a) Total-energy differences relative to the optimal reference predicted by FNEH and the FNQS as a function of $\epsilon$. The system parameters are $N_{\rm site}=36$, $N_e=12$ (1/3 filling), $V=11$meV, and $\phi=40^\circ$. (b) Energy components predicted by optimal reference, FNEH, and FNQS across $\epsilon$. FNEH energies are evaluated at intervals of $0.1$ in $\epsilon$, whereas the FNQS and optimal reference energies are evaluated at intervals of $1$. (c) Optimal weights of the FNEH basis states as a function of $\epsilon$. (d) Radially averaged pair-correlation functions predicted by the optimal reference, FNQS, and FNEH. (e) Computational time (in hours) required to scan over a number of target states using the optimal reference, FNQS, and FNEH.}
    \label{fig:epsilon}
\end{figure}

In Fig.~\ref{fig:epsilon}, we use the FNEH to study the phase transition as a function of the dielectric constant $\epsilon$. We consider a system with $N_{\rm site}=36$ and $N_e=12$ ($1/3$ filling), identical to that used in Fig.~\ref{fig:2d_phase_scan}. We fix $V=11$meV and $\phi=40^\circ$, and train an FNQS over $\epsilon\in[5,15]$. Five FNQS states at $\epsilon=5,7,9,11,$ and $13$ are selected to construct the FNEH, which is then evaluated over $\epsilon\in[6,15]$. In Fig.~\ref{fig:epsilon}a, we plot the energy difference relative to the optimal reference for both the FNQS and FNEH. While the FNQS and FNEH yield similar energies in the FL phase, FNEH substantially lowers the energy in the CDW phase, particularly for $\epsilon\in[8,10]$. This energy lowering can be understood from Fig.~\ref{fig:epsilon}b, where we compare the kinetic and Coulomb energy components for all three methods. At $\epsilon=10$, the FNQS predicts a higher $E_{\rm coul}$ than both the FNEH and optimal reference, indicating that it predicts the wrong phase. Moreover, the FNQS also predicts an incorrect transition location, placing the phase transition within $\epsilon\in[9,10]$, whereas both the FNEH and optimal reference place it within $\epsilon\in[10,11]$. Finally, the FNEH further resolves the transition more precisely at $\epsilon=10.5$.

The performance enhancement of FNEH can be understood from Fig.~\ref{fig:epsilon}c, where we plot the weight of each basis state, $|c_i|^2$, across the full parameter range. Away from the phase transition, for example around $\epsilon=7,11,$ and $15$, the dominant contribution comes from the basis state closest to the target parameter, while the coefficients evolve smoothly between neighboring basis states as $\epsilon$ varies. Near $\epsilon=10$, however, the behavior changes qualitatively. In this region, the optimal FNEH state is composed predominantly of FNQS basis states corresponding to the CDW phase. By combining information from basis states across the parameter region, FNEH is able to recover the physically appropriate CDW state and substantially lower the energy relative to the FNQS, which predicts the incorrect phase at the target parameter. More generally, although the FNQS does not locate the phase boundary accurately on its own, it still generates representative states from both sides of the transition. The subspace spanned by these states allows FNEH to accommodate the abrupt change in the optimal wavefunction across the phase boundary, thereby recovering both phases and locating the transition more accurately.

Lastly, in Fig. 4d, we examine the radially averaged pair-correlation functions predicted by the three methods at \(\epsilon=10\). Consistent with its higher Coulomb energy, the FNQS predicts an FL phase. Its dominant peaks occur near \(a_M\), \(\sqrt{3}a_M\), \(\sqrt{7}a_M\), and \(\sqrt{13}a_M\), consistent with the underlying moiré lattice. In contrast, \(g(r)\) from both FNEH and the optimal reference exhibit a dominant peak near \(\sqrt{3}a_M\), together with a plateau associated with two overlapping shells near \(3a_M\) and \(2\sqrt{3}a_M\), consistent with the CDW phase. In Fig. 4e, we compare the measured wall-clock cost of the three methods as a function of the number of target states. We assume both FNEH and FNQS use \(1.024\times10^5\) samples, while each optimal reference is trained using 1024 samples for 2000 optimization steps. The computational cost of both the optimal reference and FNQS grows linearly with the number of target states, whereas the FNEH cost remains approximately constant because the effective Hamiltonian is constructed only once. FNEH therefore becomes more computationally efficient than FNQS when the number of target states for prediction exceeds approximately 12.

\textit{Discussion---.}
In this work, we introduce the Foundation Neural Effective Hamiltonian (FNEH), which transforms FNQS from a parameter-conditioned wavefunction model into a reusable variational Hamiltonian framework over a family of couplings. By projecting each target Hamiltonian onto a compact subspace spanned by FNQS wavefunctions at selected parameter points, FNEH provides two main benefits: improved variational accuracy and reduced sampling cost. In strongly correlated moiré systems, FNEH can recover the correct phase boundary even when the FNQS misidentifies the transition. Moreover, determinant-state sampling together with an operator-resolved decomposition reduces the projected Hamiltonian to reusable parameter-independent matrices, enabling dense multidimensional parameter scans without repeated neural-network sampling. FNEH therefore integrates the transferability of foundation models with the accuracy and sampling efficiency of variational subspace methods. For future work, its performance can be further improved by using more accurate FNQS or incorporating a small number of target-optimized states through few-shot learning. Although demonstrated for continuum moiré electrons, FNEH is broadly applicable to other Hamiltonian families, including lattice fermion, spin, and molecular systems, providing a general route toward efficient foundation-model-based quantum many-body simulation.

\textit{Acknowledgement}.--- DL acknowledges support from Beijing Municipal Science and Technology Commission and Zhongguancun Science Park Administrative Committee (No. 20251090054). 

\bibliography{main}

\bibliographystyle{apsrev4-1}

\appendix 

\clearpage

\onecolumngrid
\begin{center}
	\noindent\textbf{Supplementary Material}
	\bigskip
		
	\noindent\textbf{\large{}}
\end{center}

\onecolumngrid

\section{Foundation Neural Quantum State}
The Foundation Neural Quantum State(FNQS) $|\Psi(\boldsymbol\lambda_i) \rangle$ is defined on the Hamiltonian parameter manifold $ \boldsymbol \lambda \equiv (V_0, \phi, \epsilon)$. For the FNQS architecture, we adopt a Psiformer\cite{pfau2020ab} backbone augmented with periodic features and Hamiltonian parameter embeddings. To encode electron correlations, we first derive electron feature from electron positions:
\begin{eqnarray}
\mathbf{e}_{{\rm elec},i} = \left( \sin\left(2\pi \mathbf{r}_i \mathbf{L}^{-1}\right), \cos\left(2\pi \mathbf{r}_i \mathbf{L}^{-1}\right) \right),
\end{eqnarray}
where $\mathbf{L} = (\mathbf{L}_1, \mathbf{L}_2)$ is the supercell matrix. The Hamiltonian parameters are also transformed to raw features $\mathbf{e}_{\rm h}$:
\begin{eqnarray}
\mathbf{e}_{\rm h} = \big( V_0,\, \cos\phi,\, \sin\phi,\, \frac{1}{\epsilon} \big)
\end{eqnarray}
The raw Hamiltonian features are then normalized $\widetilde{\mathbf{e}_{\rm h}} = \frac{\mathbf{e}_{\rm h} - \boldsymbol{\mu}_{\mathbf{e}_{\rm h}}}{\boldsymbol{\sigma}_{\mathbf{e}_{\rm h}}}$. The features $\mathbf{e}_{{\rm elec},i}$ and $\widetilde{\mathbf{e}_{\rm h}}$ are then embed with separate feed forward layers, and passed to the self-attention head of the transformer. 

\begin{table}[ht]
\centering
\label{tab:optim_opts}
\small
\setlength{\tabcolsep}{6pt}
\renewcommand{\arraystretch}{1.15}
\begin{tabular}{@{}l l@{}}
\toprule
\textbf{Parameter} & \textbf{Value} \\
\midrule
Optimizer        & KFAC \\
Damping          & $1\times 10^{-5}$ \\
Minimum damping  & $1\times 10^{-6}$ \\
Norm constraint  & $1\times 10^{-3}$ \\
Momentum         & $0.0$ \\
Curvature EMA    & $0.95$ \\
Estimation mode  & \texttt{fisher\_exact} \\
\bottomrule
\end{tabular}
\caption{Optimizer parameters. The learning rate is chosen as 
$\rm{lr} = \alpha \left(1 + t/\tau\right)^{-1}$, where 
$\tau = 5000$. $\alpha$ is not fixed from one run to another, and is typically around $5 \times 10^{-4}$.}
\end{table}

The FNQS is optimized over a region of the Hamiltonian parameter manifold. To stabalize training, we select a discrete set of training points $\Lambda_{\rm train}$, which leads to the following loss function:
\begin{eqnarray}
\mathcal{L} = \sum_{\boldsymbol\lambda \in \Lambda_{\mathrm{train}}} \frac{\langle \Psi(\boldsymbol\lambda) | \hat{H}(\boldsymbol\lambda) | \Psi(\boldsymbol\lambda) \rangle}{\langle \Psi(\boldsymbol\lambda) | \Psi(\boldsymbol\lambda)\rangle}
\end{eqnarray}
The gradient of $\mathcal{L}$ is then estimated stochastically via Monte Carlo sampling: 
\begin{eqnarray}
\nabla_\theta \mathcal{L} = \sum_{\boldsymbol\lambda}2\Re\{\mathbb{E}_{\boldsymbol{R} \sim p_{\boldsymbol\lambda}^2}[\frac{\partial \ln{|\Psi(\boldsymbol\lambda, \boldsymbol{R})|}}{\partial \theta}E_{l}(\boldsymbol\lambda, \boldsymbol{R})]\}
\end{eqnarray}
where $p_{\boldsymbol\lambda}^2 \propto |\Psi(\boldsymbol\lambda)|^2$ is the probability distribution of the foundation wavefunction conditioned at $\lambda$, $E_{l}(\boldsymbol\lambda, \boldsymbol{R}) = \frac{\hat{H} \Psi}{\Psi}(\boldsymbol\lambda, \boldsymbol{R})$ is the local energy. 

We employ different training strategies to stabilize the FNQS optimization and avoids local minima. For the FNQS used in Fig.~\ref{fig:energy_bench} and Fig.~\ref{fig:2d_phase_scan}, we initialize training using two parameter points located on opposite sides of the phase boundary, and subsequently increase the number of training points as optimization proceeds. For the FNQS used in Fig.~\ref{fig:epsilon}, we perturb the training parameter at each optimization step with Gaussian noise, using a standard deviation of $0.5$ in $\epsilon$. For all our trainings, we use the Kronecker-factored Approximate Curvature (KFAC)\cite{kfac-jax2022github} optimizer implemented on Jax\cite{jax2018github} to precondition the gradient. The optimizer details can be find in Table.~\ref{tab:optim_opts}.

\section{Benchmarking methods}
Benchmarking methods used in this work, including band projected exact diagonalization (BPED) and Hartree-Fock (HF), works under the second quantization formulism. The single particle Hamiltonian in second quantization is ${\hat H}_0 = \int d\boldsymbol{r} \psi^\dagger(\boldsymbol{r}) \hat{h} \psi(\boldsymbol{r})$, where $\hat{h}$ is the single particle part of Eq.~\ref{EQ:Main_Hamil}. For arbitrary momentum in reciprocal space, its mesh component can be written as $[\boldsymbol{q}] = \boldsymbol{q} - (m_{1} \boldsymbol{b}_1 + m_{2} \boldsymbol{b}_2)
$ ($m_{i} \in \mathbb{Z}$), where $\boldsymbol{b}_1$ and $\boldsymbol{b}_2$ are the basis of reciprocal lattice (We choose ${\boldsymbol{b}_1} = \boldsymbol{g}_1$ and ${\boldsymbol{b}_2} = \boldsymbol{g}_3$ from here onwards). Similarly, the reciprocal lattice component of $\boldsymbol{q}$ can be written as $\boldsymbol{g}_{\boldsymbol{q}} = m_{1} \boldsymbol{b}_1 + m_{2} \boldsymbol{b}_2$. We denote $\boldsymbol{k}$ as an arbitrary momentum in the First Brillouin Zone (FBZ), and $\boldsymbol{g}$ as an arbitrary reciprocal lattice vector. Fourier expanding $\psi^\dagger(\boldsymbol{r})$ and $\psi^(\boldsymbol{r})$ enables diagonalization of ${\hat H}_0$ in $\boldsymbol{k} + \boldsymbol{g}$ basis, yielding Bloch states and band energies:
\begin{eqnarray}
\ket{u_{\boldsymbol{k}, n}} = \sum_{\boldsymbol{g}, l}u_{\boldsymbol{k},l}^{(n)}(\boldsymbol{g}) \ket{\boldsymbol{k} + \boldsymbol{g},l} ~~~~~\text{with}~~~~~ {\hat H}_{0}(\boldsymbol{k}) \ket{u_{\boldsymbol{k}, n}} = \epsilon_{\boldsymbol{k}, n} \ket{u_{\boldsymbol{k}, n}}
\end{eqnarray}
where $n$ is band index. To construct the many-body Hamiltonian, we define Bloch state creation and annihilation operator ${\hat c}_{\boldsymbol{k}, n}^\dagger$ and ${\hat c}_{\boldsymbol{k}, n}$. The many-body wavefunction is effectively projected onto a bitstring basis $\ket{\boldsymbol x} = \ket{x_1, x_2, \dots, x_{N_{q}} }$, where each $x_j \in \{0,1\}$ encodes the occupation of the Bloch state $\ket{u_{\boldsymbol{k}_j, n_j}}$. $N_{s} = N_{k}\times N_{b}$ is the total number of single-particle Bloch states, and $N_{b}$ is the total number of bands. Numerically, we restrict $\ket{u_{\boldsymbol{k}, n}}$ such that $ \epsilon_{\boldsymbol{k}, n} \in N_b \text{ lowest eigenvalues of }\hat{H}_{\uparrow}(\boldsymbol{k} )$. We choose $N_b = 1$ for exact diagonalization, and $N_b = 6$ for HF calculations. The band projection leads to the following Hamiltonian: 
\begin{eqnarray}\label{BP_ham}
{\hat H}_{\rm BP} = \sum_{i} \epsilon_{\boldsymbol{k}_i, n_i}c^\dagger_{\boldsymbol{k}_i, n_i} c_{\boldsymbol{k}_i, n_i} + \frac{1}{2} \sum_{i,j,k,l} {\hat V}_{i,j,k,l} c^\dagger_{\boldsymbol{k}_i, n_i}c^\dagger_{\boldsymbol{k}_j, n_j} c_{\boldsymbol{k}_l, n_l}c_{\boldsymbol{k}_k, n_k}
\end{eqnarray}
where $i,j,k,l$ are bit indices. The Coulomb tensor takes the form of:
\begin{eqnarray}\label{coulomb_tensor}
{\hat V}_{i,j,k,l} &=& \frac{1}{A}\sum_{\boldsymbol q} V({\boldsymbol q})\bra{u_{{\boldsymbol k}_i, n_i}}e^{i{\boldsymbol q}{\boldsymbol r}_{1}}\ket{u_{{\boldsymbol k}_k, n_k}} \bra{u_{{\boldsymbol k}_j,n_j}}e^{-i{\boldsymbol q}{\boldsymbol r}_{2}}\ket{u_{{\boldsymbol k}_l,n_l}}
\end{eqnarray}
where $V({\boldsymbol q}) = \frac{2\pi e^2}{\epsilon |{\boldsymbol q}|}$. $A = N_{k} \cdot|{\boldsymbol b}_1 \times {\boldsymbol b}_2|$ is the supercell area. We use two different supercells in this work. For $N_{\rm site}=36$, the supercell matrix is $M=[[6,0],[0,6]]$, while for $N_{\rm site}=27$, the supercell matrix is $M=[[3,3],[3,-6]]$. $F(\boldsymbol{k}_1,\boldsymbol{k}_2, n_1, n_2, \boldsymbol{g}) = \sum_{\boldsymbol{g}^\prime, l}u_{\boldsymbol{k}_1,l}^{(n_1)\ast}(\boldsymbol{g}^\prime+ \boldsymbol{g})u_{\boldsymbol{k}_2,l}^{(n_2)}(\boldsymbol{g}^\prime) $ is the form factor. Numerically, ${\boldsymbol q}$ should be summed over all possible momentum transfer allowed the $\boldsymbol{k} + \boldsymbol{g}$ basis. Directly diagonalizing Eq.~\ref{BP_ham} gives the BPED solution.  

The HF formalism can also be applied to Eq.~\ref{BP_ham}. We adopt the formalism from Ref.~\cite{dai2024strong}. Define coefficient matrix $\ket{\phi_i} = C_{\mu i} \ket{u_{\boldsymbol{k}_{\mu}, n_{\mu}}}$, the Hartree-Fock wavefunction can be written as $\Ket{\Psi}_{\rm HF} = {\rm Det} [\phi_i(\mathbf{r}_j)]$. The matrix elements in the one-particle basis can be written as:
\begin{eqnarray}
{h}_{\mu\nu} = \langle \mu|{\hat H}_0 |\nu \rangle = \delta_{\mu \nu} \epsilon_{\boldsymbol{k}_{\mu}, n_{\mu}} ~,\quad v_{\mu \nu \kappa \lambda} = \langle \mu \nu | {\hat U} |\kappa \lambda \rangle = {\hat V}_{\mu,\nu,\kappa,\lambda}
\end{eqnarray}
The HF energy can be written as:
\begin{eqnarray}
E = \Big[ {h}_{\mu\nu} + \frac{1}{2}(v_{\mu \kappa \nu \lambda} - v_{\mu \kappa \lambda \nu} ) D_{\kappa \lambda} \Big] D_{\mu \nu}
\end{eqnarray}
where $D_{\mu \nu} = C_{\mu i}^\ast C_{\nu i}$ is the one-particle reduced density matrix. The analytical gradient of energy with respect to coefficient matrix is then:
\begin{eqnarray}
\frac{\partial E}{\partial C_{\mu \nu}^\ast} = [(\mathbf{I} - \mathbf{C}\mathbf{S}^{-1} \mathbf{C}^\dagger)\mathbf{F}\mathbf{C}\mathbf{S}^{-1}]_{\mu \nu}
\end{eqnarray}
where $\mathbf{S} = \mathbf{C}^\dagger \mathbf{C}$ is the overlap matrix. $\mathbf{F}$ is the Fock matrix, which can be written as: $F_{\mu \nu}[D] =  {h}_{\mu\nu} + \frac{1}{2}(v_{\mu \kappa \nu \lambda} - v_{\mu \kappa \lambda \nu} ) D_{\kappa \lambda}$. We use the Broyden–Fletcher–Goldfarb–Shanno (BFGS) minimizer with analytical gradient implemented on \texttt{jax.scipy} to find the minimal Hartree-Fock energy. The gradient norm tolerance is set to $1 \times 10^{-6}$ to ensure convergence.

\section{Determinant State Calculations}

\subsection{Proof of Eq.~\ref{EQ:det_hamil}}
Previous work has discussed the determinant state sampling in wavefunction subspace~\cite{hendry2025grassmann, pfau2024accurate, kahn2026variational}. Here, we provide a detailed proof of Eq.~\ref{EQ:det_hamil}. For simplicity, denote $\psi_j(\mathbf R):=\Psi(\boldsymbol\lambda_j,\mathbf R),\, j=1,\ldots,N_{\rm state},$ and define the matrix $\mathbf\Psi(\{\mathbf R\})_{ij} = \psi_j(\mathbf R_i).$

Using the Andréief identity, for a set of single-coordinate functions $\{f_i(\boldsymbol r)\}_{i=1}^{N_{\rm state}}$ and $\{g_j(\boldsymbol r)\}_{j=1}^{N_{\rm state}}$,
\begin{equation}
    \int\det[f_j(\boldsymbol R_i)]\det[g_j(\boldsymbol R_i)]\prod_{j=1}^{N_{\rm state}} d\boldsymbol R_i=
    N_{\rm state}!\det\left[\int f_i(\boldsymbol r)g_j(\boldsymbol r)d\boldsymbol r\right].
\end{equation}
with $f_j^*(\mathbf r)=g_j(\mathbf r)=\psi_j(\mathbf r),$ one obtains
\begin{equation}
    Z=\int|\det\mathbf\Psi|^2\,d\{\mathbf R\}=N_{\rm state}!\det(S),
\end{equation}
where $d\{\boldsymbol R\}=\prod_{j=1}^{N_{\rm state}}d\boldsymbol R_j.$ and $S_{ij}=\langle\psi_i|\psi_j\rangle.$

Now consider the local Hamiltonian estimator
\begin{equation}
    A_k(\{\mathbf R\}) = \boldsymbol{\Psi}^{-1}[{\hat h}^{(k)}_{\rm det} \boldsymbol{\Psi}].
\end{equation}
Its $(i,j)$-th entry is
\begin{equation}
    (A_k)_{ij}=\sum_{l=1}^{N_{\rm state}}(\mathbf\Psi^{-1})_{il}(\hat h_k\psi_j)(\mathbf R_l).
\end{equation}
Using the adjugate formula $\mathbf\Psi^{-1}=\frac{\operatorname{adj}(\mathbf\Psi)}{\det\mathbf\Psi},$ we have
\begin{equation}
    |\det\mathbf\Psi|^2(A_k)_{ij}=(\det\mathbf\Psi)^*\sum_{l}(\operatorname{adj}\mathbf\Psi)_{il}(\hat h_k\psi_j)(\mathbf R_l).
\end{equation}
By the Laplace expansion along the $i$-th column, the summation on the right-hand side is exactly the determinant obtained by replacing the $i$-th column of $\mathbf\Psi$ with $(\hat h_k\psi_j(\mathbf R_1),\ldots,\hat h_k\psi_j(\mathbf R_{N_{\rm state}}))^T$.
Denoting this matrix by $\mathbf\Psi_k^{(i,j)}$, we obtain
\begin{equation}
    |\det\mathbf\Psi|^2(A_k)_{ij}=(\det\mathbf\Psi)^*\det(\mathbf\Psi_k^{(i,j)}).
\end{equation}
Applying the Andréief identity once again to this expression yields
\begin{equation}
    \int|\det\mathbf\Psi|^2(A_k)_{ij}\,d\{\mathbf R\} = N_{\rm state}!\,\det(S_k^{(i,j)}),
\end{equation}
where $S_k^{(i,j)}$ denotes the overlap matrix whose $i$-th column is replaced by the $j$-th column of matrix $\tilde h_k=\langle\psi_i|\hat h_k|\psi_j\rangle.$

Finally, by Cramer's rule,
\begin{equation}
    \det(S_k^{(i,j)})=\det(S)(S^{-1}\tilde h_k)_{ij},
\end{equation}
which yields
\begin{equation}
    \int|\det\mathbf\Psi|^2(A_k)_{ij}\,d\{\mathbf R\}=N_{\rm state}!\det(S)(S^{-1}\tilde h_k)_{ij}.
\end{equation}

Dividing both sides by the normalization constant
\(Z=N_{\rm state}!\det(S)\)
gives
\begin{equation}
    \mathbb E_{\{\textbf{R}\}\sim p_{\rm det}}[(A_k)_{ij}] = (S^{-1}\tilde h_k)_{ij} = (\hat A_k)_{ij}.
\end{equation}

Since this identity holds for every matrix element, we conclude
\begin{equation}
\mathbb E_{\{\mathbf R\}\sim p_{\rm det}}\left[\mathbf\Psi^{-1}[{\hat h}^{(k)}_{\rm det}\mathbf\Psi]\right] = S^{-1}\tilde h_k = \hat A_k,
\end{equation}
which proves Eq.~\ref{EQ:det_hamil}.

\end{document}